\documentstyle[11pt,newpasp,twoside,epsf]{article}
\def\edcomment#1{\iffalse\marginpar{\raggedright\sl#1\/}\else\relax\fi}
\reversemarginpar

\begin{document}
\title{{\it HST} Observations of the Nuclear Regions of the Toomre Sequence of 
Merging Galaxies}
 \author{S. Laine, R. P. van der Marel and T. B\"{o}ker}
\affil{STScI, 3700 San Martin Drive, Baltimore, MD 21218, USA}
\author{J. C. Mihos}
\affil{Department of Astronomy, Case Western Reserve University, 10900 Euclid
Avenue,Cleveland, OH 44106, USA}
\author{J. E. Hibbard}
\affil{NRAO, 520 Edgemont Road, Charlottesville, VA 22903, USA}
\author{A. I. Zabludoff}
\affil{Department of Astronomy, Univ. of Arizona, 933 North Cherry Avenue, 
Tucson, AZ 85721, USA}

\begin{abstract} We present first results from an {\it HST} WFPC2 imaging and
STIS spectroscopy program to investigate the structural and star forming
properties in the nuclear regions of the Toomre Sequence of merging galaxies.
Here we discuss $V$-band, $I$-band and H$\alpha$ images of the
nuclei. We comment briefly on the connection between the nuclear morphology of
the ionized gas and the merger stage. 
\end{abstract}

\section{Introduction}

Observational studies of merging galaxies have shown a clear link between
interactions and nuclear activity, while dynamical models have revealed the
physics driving the radial inflows of gas. Unfortunately, neither ground-based
observations nor current simulations are able to accurately follow the
evolution of the gas once it reaches the inner few hundred parsecs of the
galaxy. Whether the gas flow ``hangs up'' in the inner few hundred parsecs and
forms stars, or continues to flow inward towards a putative AGN has a strong
impact on the luminosity and evolution of the merger. Numerical simulations
(e.g.,~Mihos \& Hernquist 1996; Barnes \& Hernquist 1996) suggest that the
nuclear gas dynamics will depend on the structure of the host galaxies and on
the dynamical stage of the merger.

\begin{figure}
\plotfiddle{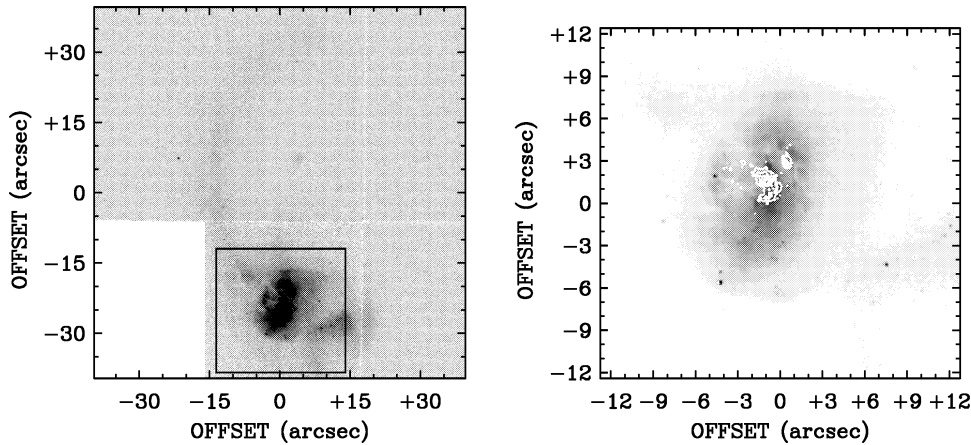}{5cm}{270}{50}{50}{-195}{225}
\caption{$V$-band image of NGC 2623 (left) with a rectangle indicating the
magnified area on the right which shows the ionized gas disk in H$\alpha$ 
emission (white contours).}
\end{figure}

To study the evolution of the nuclei in the course of a merger in greater
detail, we have acquired {\it HST} observations of the galactic nuclei in the
Toomre Sequence of merging galaxies (Toomre 1977). This is a sequence of eleven
systems in progressively more advanced stages of merging, from galaxies in
early stages of merging (NGC 4038/9 and NGC 4676) to late stage remnants  (NGC
3921 and NGC 7252).

\section{Data}

We have taken 320 sec CR-SPLIT exposures of each galaxy nucleus using the WFPC2
F555W ($V$) and F814W ($I$) filters. The galaxy nuclei were placed on the
Planetary Camera chip (0.046 arcsec pixels) to achieve the highest spatial
resolution. We also took longer, 1200 sec exposures in the H$\alpha$ + [N II]
lines, mostly through the Linear Ramp Filter (which has $\approx$ 10$''$
$\times$ 10$''$ arcsec useful field of view). The latter observations allow us
to examine the ionized gas distribution.

We will also acquire {\it HST} STIS spectroscopic data of the same galaxy
nuclei. The G750M grating will be used to obtain H$\alpha$ spectroscopy of the
detected ionized gas disks. Three parallel slit positions will be used along
the major axis of the gas disks. These two-dimensional mappings will allow us
to discriminate between organized and chaotic motions in the gas (e.g. Mihos \&
Bothun 1998). We will also use the G430L grating to determine the stellar
populations in the nuclear regions by comparing the observations to population
synthesis models (e.g.,~Bruzual \& Charlot 1996; Leitherer et al. 1999).

\section{Project Goals}

With the {\it HST} data we will be in a position to address several important
questions regarding the structure and evolution of merger remnants: 

\begin{enumerate}

\item The morphology of the gas distribution in the innermost kiloparsec of
merging galaxies is poorly known. Does the gas settle in a resonance ring
(e.g. Hernquist \& Mihos 1995) or does it continue to flow all the way to the
nucleus, forming a nuclear disk (e.g.,~Jaffe et al. 1999; van der Marel \& van
den Bosch 1998)?
\looseness=-2

\item Since the morphology of the circumnuclear gas distribution is often
determined by dynamical resonances, we will look for bars near the nuclei of
merging galaxies. Because the resonance locations are partly determined by the
rotation curve, we will examine the shape of the rotation curve near the
nuclei.

\begin{figure}
\plotfiddle{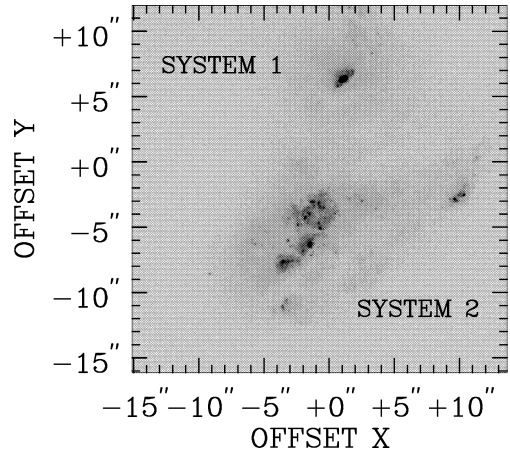}{5.0cm}{0}{32}{32}{-130}{-45}
\caption{NGC 7592, showing an example of a point-like and a hidden nucleus.}
\end{figure}

\item We will perform a systematic observational study to determine how the
circumnuclear gas kinematics change as the merger advances. Specifically, we
will assess whether the gas kinematics become more ordered or chaotic as the
merger progresses.

\item Since large amounts of gas are speculated to flow into the nuclear region
as the merger progresses, we will look for signatures of supermassive black
holes in the nuclei and AGN activity.

\item We will inspect whether the stellar population resulting from a
circumnuclear starburst follows the R$^{1/4}$ law. We will also determine at
which stage of the merger star formation activity peaks, and also the size of
radial gradients of color and line strength in the nuclear starburst
population. \end{enumerate}

\section{Early Results}

\begin{figure}
\plotfiddle{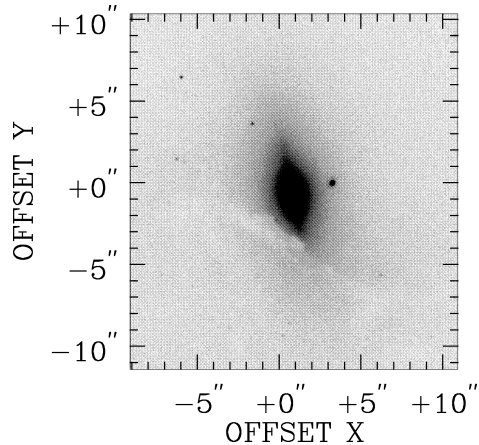}{5.0cm}{0}{30}{30}{-120}{-42}
\caption{An {\it HST} $I$-band image of NGC 6622, showing a bar-like feature
around the nucleus.}
\end{figure}

Example WFPC2 images are shown in Figures 1--3. Our preliminary analysis of
these data has produced the following results:

\begin{enumerate}

\item We have detected nuclear H$\alpha$ gas disks in six of eleven systems
(NGC 2623; Fig. 1, 3256, 3509A, 3921, 4038, 7252). It is perhaps significant that all
the four latest stage mergers have nuclear gas disks.

\item All the five systems in our sample which consist of two clearly
identified separate galaxies have one galaxy with a bright, point-like
nucleus, whereas the other galaxy has lost the identity of its nucleus (e.g.,
NGC 7592A; Figure 2). This dichotomy may come from inclination/obscuration
effects or differences in progenitor morphological types. 

\item Bar-like configurations are seen in a few galaxies (e.g., NGC 6622;
Fig. 3). Kinematical data will reveal whether they are highly inclined disks
or genuine stellar bars. \end{enumerate} 

\acknowledgments 
Support for this work was provided by NASA through grant
for project GO-8669 (P.I. van der Marel), awarded by the STScI which is
operated by AURA, Inc., under NASA contract NAS5-26555.

\end{document}